\documentclass[aps, prl, preprint,groupedaddress]{revtex4-1}

\draft 
\usepackage{graphicx}
\usepackage{caption}
\usepackage{color}

\begin{document}


\title{Contact Mechanics of a Small Icosahedral Virus} 



\author{Cheng Zeng}
\affiliation{Department of Chemistry, Indiana University, Bloomington, IN 47405, U.S.A.}

\author{Mercedes Hernando-P\'{e}rez}
\affiliation{Department of Chemistry, Indiana University, Bloomington, IN 47405, U.S.A.}

\author{Xiang Ma}
\affiliation{Department of Chemistry, Idaho State University, Pocatello, ID 83209, U.S.A.}

\author{Paul van der Schoot}
\altaffiliation{Institute for Theoretical Physics, Utrecht University, Leuvenlaan 4, 3584 CE Utrecht, The Netherlands}
\affiliation{Department of Applied Physics, Eindhoven University of Technology, P.O. Box 513, 5600 MB Eindhoven, The Netherlands.}

\author{Roya Zandi}
\affiliation{Department of Physics and Astronomy, University of California at Riverside, 900 University Ave.
Riverside, CA 92521, U.S.A.}

\author{Bogdan Dragnea}
\email[]{dragnea@indiana.edu}
\affiliation{Department of Chemistry, Indiana University, Bloomington, IN 47405, U.S.A.}


\date{\today}

\begin{abstract}

Virus binding to a surface results at least locally, at the contact area, in stress and potential structural perturbation of the virus cage. Here we address the question of the role of substrate-induced deformation in the overall virus mechanical response to the adsorption event. This question may be especially important for the broad category of viruses that have their shells stabilized by weak, non-covalent interactions.   We utilize atomic force microscopy to measure the height change distributions of the brome mosaic virus upon adsorption from liquid on atomically flat substrates and present a continuum model which captures well the behavior. Height data fitting according the model provides, without recourse to indentation, estimates of virus elastic properties and of the interfacial energy.

\end{abstract}


\maketitle 


The problem of how adhesion of a deformable object to a surface is driven by interfacial energy and opposed by elasticity is at the center of modern contact mechanics\cite{Maugis1999}.  Cell membranes are naturally impermeable to virus particles. For viruses to cross plasma, endosomal, or nuclear membranes, the virus-cell interface has to change drastically after virus adsorption. This is often done in a system-specific manner.  Nevertheless, before specific transformations to take place, virus particles must stick at the apical cell surface via generic interactions, e.g., hydrophobic or electrostatic\cite{Mercer2010}. Could this initial, random binding event already perturb the mechanochemistry of the virus particle in a way that would prime it for the next sequence in the entry process?  Gao et al. have suggested a model for the clathrin-independent endocytosis mechanism by which interactions between ligands fixed on the particle surface and free receptors on the plasma membrane would result in bringing more of the membrane into contact with the particle, which in turn would  lead to the particle being eventually engulfed by the plasma membrane\cite{Gao2005}. This receptor-mediated wrapping mechanism model was revisited by Yi et al. who allowed particles to deform under the influence of adhesion to the flexible membrane surface and pointed out the possibility of a strong effect of the elastic deformation of particles on their cellular uptake\cite{Yi2011}. Furthermore, more recent experimental studies provided indication that, at least in certain cases, virus stiffness may regulate entry\cite{Pang2013}.

In contact mechanics of small soft-material particles, solid surface tension dominates elasticity\cite{Style2013}. Yet, this aspect has not been considered so far in approaches to mechanical measurements of viruses by atomic force microscopy (AFM) nanoindentation. Here we report on a case study aiming to determine how virus mechanics responds to virus adsorption to a surface. We find that a small icosahedral plant virus, the brome mosaic virus (BMV), will bind to atomically flat surfaces predominantly in one orientation, and that in order to achieve this preferred orientation it will deform, mainly at the contact interface. Moreover, in indentation experiments, the spring constant of the virus was independent of substrate-induced deformation. In other words, local stresses due to surface binding and distortion do not seem to propagate to the top, where the measurement is done. Interestingly, this would also be expected within the framework of  thin shell theory\cite{buenemann2007}. Furthermore, with the aid of an elastic model, we show how the distribution of particle heights on the substrate informs on the magnitudes of elastic moduli and of the contact surface energy, without recourse to indentation experiments.

Viruses are obligated biological systems much smaller in size than cells, but still composed of hundreds to tens of thousands of molecules working together. A complete understanding of their dynamic behavior requires a unifying framework including contributions from scale-dependent and scale-independent phenomena\cite{Phillips06}. In recent years, studies of virus mechanics under the influence of an external perturbation have begun to shed light on how energy flows between the different degrees of freedom describing these complex molecular assemblies. Osmotic pressure assays have provided important clues on how chemical energy is transformed into mechanical energy for phage genome injection\cite{Gelbart:2008wt} and single molecule pulling experiments with optical tweezers have helped elucidating the mechanisms of phage genome packaging\cite{Smith:2001cz, Purohit2003, Smith2011}. 

In the category of \emph{in singulo} methods based on mechanical force application, AFM indentation\cite{Michel2006} has  allowed the measurement of virus and protein cage deformation under uniaxial load\cite{Roos2010a, Carrasco2011}, and of the relationship between virus mechanics and chemistry, which includes contributions from environmental factors such as pH or hydration\cite{Wilts2015}, and from the nucleic acid cargo\cite{Michel2006, Carrasco2006a, Vaughan2014, Ahadi2013}. For sufficient imaging resolution and to perform reproducible  indentation experiments,  particles have to be  immobilized strongly enough to resist lateral forces exerted by the AFM probe\cite{Baclayon2010}. Compared with contact modes, non-contact modes\cite{Hansma1994,Moreno-Herrero2003, Martinez-Martin2012} are generally considered as being the least intrusive. Even then, while mean forces during imaging are usually below 0.1 nN,  peak force estimates in ``tapping" mode can exceed 0.1 nN (albeit for only $\sim$1 ms per pixel)\cite{Xu2008}. Such forces require either virus immobilization in a crystalline lattice\cite{Kuznetsov1997} or, when single virus measurements are sought, strong adhesion forces between virus and substrate. This is why substrates are usually prepared by coating with molecules imparting a hydrophobic or charged character to the surface \cite{Muller1997, Roos2011a}. Thus, when the virus binds to the surface through, say, hydrophobic interactions, an equilibrium is established between virus-substrate adhesive interactions and the cohesive interactions of the virus.  Adhesion-related deformation was observed before\cite{Knez2004}, but very little is actually known about this equilibrium. How does the balance between adhesion and mechanical stresses affect particle shape?  How large is the adhesion area at equilibrium?  What is the magnitude of surface energy? Does surface adhesion result in local structural perturbations that propagate through the virus lattice up to the top, at the indentation area? This study takes on the task of addressing these questions on one of the most-studied virus systems adsorbed on chemically well-defined, atomically flat substrates. 
 
BMV was the first virus to be imaged by AFM at molecular resolution\cite{Kuznetsov2001}. It is an established model\cite{Kao2000} for small (+) single--stranded RNA icosahedral viruses, the most plentiful viruses on this planet\cite{Flint2009}. BMV has a non-enveloped capsid formed from 180 copies of the same coat protein (CP), organized in a T=3 lattice with an average outer diameter of 284 \AA\cite{Wang2014}. The outer surface of the BMV capsid is studded with hydrophobic patches surrounded by polar residues (Fig. SI-1) and thus BMV readily adsorbs on both hydrophobic surfaces and polar surfaces.  

In this work, we study the distributions of maximum particle heights measured by AC-mode AFM on two substrate materials that readily yield atomically flat surfaces: highly-oriented pyrolytic graphite (HOPG), and mica. The idea is that adhesion forces will tend to maximize the contact area by locally flattening the virus at the contact point. Assuming that, for small perturbations, the virus particle behaves approximately as an elastic shell\cite{Roos2010a}, an increase in contact area can be accomplished at the energetic cost of bending the shell and of forming a rim, defined as the locus where the fluid, the substrate, and the shell outer surface meet. As a result, the maximum height of the virus over the surface support would change upon adsorption. Since measuring height is done relative to the substrate, it is beneficial to chemically homogeneous, atomically flat substrates for this work,  as opposed to functionalized etched glass substrates customarily used in indentation experiments, which have higher local roughness\cite{Roos2011a}. 

Height measurements can be affected not only by substrate roughness, but also by virus shell anisotropy. For BMV, the root-mean-square deviation (rmsd) from a spherical surface is $\sim$ 20\AA\cite{Lucas2002}. Since the measurement is made top-down, it is important to record the orientation of the virus particle relative to the substrate. Imaging at the experimental conditions reported here (see Supporting Information)  leads to sufficient lateral resolution (Fig. SI-2) to distinguish not only broad icosahedral symmetry features, but individual capsomers on the virus surface, Fig. 1. In these conditions, we find a clear orientational bias, on both substrates. The most frequent  orientation is with a three-fold axis normal to the substrate ($\sim70\%$ from a total of 38 particles which had enough resolution to be unambiguously analyzed). Note that, if particles were adsorbed with random orientation, one would expect the three-fold axis orientation to be observed significantly less often. Early work on cowpea chlorotic mottle virus done on KOH etched glass and silanized glass found random capsid orientations, in contrast with our findings\cite{Michel2006}. The difference is likely coming from the fact that etched glass is rough and chemically heterogeneous. As a consequence, particles may bind upon landing with an enhanced initial contact area, and hence with strong initial adhesion and without subsequent reorientation. The situation is likely different on atomically flat, chemically homogenous surfaces, where an initial small contact would require reorientation to avoid desorption.  Orientational selection could come from the most exposed areas on the virus surface having a pronounced hydrophobic character and affinity for nonpolar surfaces such as HOPG, Fig. SI-1. Moreover, anionic residue patches bordering these areas may bind to divalent cations (such as Mg(II) present in buffer solution) and adsorbed on the mica surface\cite{Hansma1996}. 

Histograms of BMV maximum heights on HOPG and mica are presented in Fig. 2 a). A small (5 \AA) correction to the apparent height values was made  to account for compression under the imaging force (Supporting Information)\cite{Roos2011a}. There are two notable observationss: (1) On both substrates, height histograms are peaked at values 1.5-2.0 nm below the nominal 28.4 nm for BMV. (2) Histograms are asymmetric, with the longer tail extending towards lower heights, while the greater heights wing ends abruptly in the vicinity of the nominal BMV diameter. Height distribution peak position and peak width depend on the substrate, suggesting a chemical effect.

Lower heights than the nominal diameter suggest particle deformation upon adhesion, Fig. 2 b). BMV particles have elastic constants of $\sim$ 0.2 N/m. The compression force that would have to act on the virus to obtain a deformation associated with the observed drop in height of  $\sim$ 2 nm, is $\sim$ 400 pN. This value gives an order of magnitude estimate of substrate-induced interactions at work. 

An analytical model was setup that captures in a formal, albeit heuristic way,  the interplay between elastic properties, capsid deformation, and adhesion. The model is inspired by the Helfrich treatment of the elastic properties of lipid bilayers\cite{Helfrich1973}, but with significant differences as a viral shell is a very different object than a lipid vesicle. A specific assumption is made that, upon landing on a surface, adhesion can increase by local deformation and formation of a flat contact area (base) with circular symmetry, Fig. SI-3. In other words, there is a sharp boundary or fracture between the flat surface base and the spherical cap, in solution. We opted for this geometry instead of the one  assuming continuous deformation of membrane vesicles adsorbed on a surface\cite{Seifert1990} because due to the discrete nature of shell subunits, line fracture rather than continuous deformation is a reasonable assumption. Moreover, a continuously-deformed particle should become pre-stressed by adsorption and presumably, show changes in apparent stiffness, a situation which, as we will see later, we do not observe. 

Stretching would imply deformation of the proteins and/or increase in capsomeric surface-to-surface distances. Both processes are expensive, the latter on account of the short-ranged nature of the interactions\cite{buenemann2007}. Moreover, attempts to fit the data including stretching/compression of the surface area either failed or showed that the contribution of stretching energy is negligible. Area conservation upon deformation is thus assumed, which leads to a relationship between height and the cap radius:
\begin{equation}
a = \sqrt{\frac{4r_0^2-h^2}{2}}
\end{equation} 
where $r_0$ is the initial particle radius, and $h$ is the height on the surface after binding and deformation (Fig. SI-3).
The spherical cap radius then obeys (Fig. SI-3):
\begin{equation}
r = \frac{4r_0^2+h^2}{4h} 
\end{equation}
The total energy is partitioned into contributions from the bending and Gauss energies, as well as a surface energy associated with the contact area, and a line or rim energy associated with the contact perimeter. The total energy is (see Supporting Information):

\begin{equation}
F=\frac{1}{2}\kappa \left (  \frac{2}{r}-\frac{2}{r_0} \right )^2 2 \pi r h + 2 \pi \kappa_G \frac{h}{r} -\gamma  \pi a^2 + \tau  2 \pi a
\label{eq:F}
\end{equation}
where: $\kappa$ is the bending modulus, $\kappa_G$ is the Gauss modulus,  $\gamma$ is the  surface energy, and $\tau$ is the rim energy. 

We emphasize that we attempted to fit the data without the Gaussian term and in the presence of stretching energy, see below and the SI. The expression given in Eq.~\ref{eq:F} is the simplest equation with which we are able to fit the experimental data

Within the thin shell approximation, the Gauss and bending moduli are related via Poisson's ratio\cite{Landau1984}: $\kappa_G = \kappa(\nu-1)$. For small icosahedral ssRNA viruses, $\nu \approx 0.3 - 0.4$\cite{Ahadi2013, Michel2006, Gibbons:2008kv}. Here we take the value $\nu = 0.3$. The free energy change upon adsorption can be then written as a sole function of the reduced height, $H = h/ 2 r_0$. Parameters $\kappa$, $\kappa_G$, $\gamma$, and $\tau$ can be then in principle found from fitting experimental data with a Boltzmann distribution derived from the free energy as a function of $H$ (eq.~\ref{eq:F}).

Note that the Gauss term would have a vanishing contribution on a continuous surface topologically equivalent to a sphere\cite{Seifert1990}. In our case the surface is not differentiable everywhere, and hence the Gauss term does contribute. Interestingly, if we remove the Gauss term, we obtain a substrate-dependent bending modulus, which should have been a property determined by the nature of the virus rather than the underlying surface. Furthermore, unreasonably high values for the bending modulus were also observed (see Supporting Information).  If we keep the Gauss term, then our data could be fitted using same values for the bending modulus on different substrates. The fitting results and parameters for these conditions are summarized in Fig. 3 and Table 1. As discussed in the following, parameter values agree well with those previously reported  by other methods.

Since the bending of a shell involves compression of the inner surface and extension of the outer surface, the bending modulus, $\kappa$, is related to the stretching modulus, $\kappa_s$ through: 
\begin{equation}
\kappa=\kappa_s\cdot\frac{w^2}{\alpha}
\label{eq:ks}
\end{equation}
where: $w$ is the shell thickness, and $\alpha = $ 12, 24, or 48 depending on the shell model (12 for a uniform plate\cite{Landau1984}, 24 for a polymer brush\cite{Rawicz2000}, 48 for a two-leaflet structure\cite{Boal2002}). For virus capsids, $\alpha =$ 12 has been previously used\cite{gibbons2007nonlinear,cuellar2010size,roos2010b}, which in our case leads to $\kappa_s\approx 43\ k_BT/nm^2$. In an examination of the low-frequency modes of a very similar virus to BMV, the chlorotic cowpea mosaic virus (CCMV), May \emph{et al.} calculated, in the context of a spherical harmonic basis set, $\kappa_s$ values for the $l=0$ and $l=1$ modes at  $81\ k_BT/nm^2$ and $60\  k_BT/nm^2$, respectively\cite{May:2011jx}. Note that while in AFM indentation experiments the $l=1$ is the dominating mode, both $l = 0$ and $l = 1$ modes are likely to be required in order to describe deformation in our case. Thus, estimates for the bending modulus from the particle height data lead to comparable values with those previously reported from similar systems.

\begin{table}
\caption{\label{Table 1} Fit parameter values for data in Fig. 3}
\begin{tabular}{ | l | c | c | c |}
\hline
Parameters: & $\kappa$  & $\gamma$  & $\tau$ \\
			& $(k_BT)$	& $(k_BT/nm^2)$	& $(k_BT/nm)$ \\ \hline
HOPG  & 32 & 0.16  &  0.09 \\ \hline
Mica & 32 & 0.10 & -0.30 \\
\hline

\end{tabular}
\end{table}

From the relation between the particle height, initial radius and radius of the flat part of the adsorbed virus, we can find the base area that corresponds to the most probable particle height. The base radius for HOPG is $\approx$ 9 nm and the corresponding base area is $\approx$ 250 nm$^2$. Creating the base lowers the particle energy by $\approx 40\ k_BT$ with a rim contribution of $\approx \ 5\ k_BT$. Note that different contributions dominate at different height ranges. For instance, the rim contribution dominates when the contact area is small. Populations at $ h \approx 2r_0$ are determined by the magnitude (and sign) of $\tau$. More specifically, a barrier to adsorption would occur if $\tau$ is positive  (see Fig. SI-3). Such seems to be the case on HOPG, but not on mica (see Table 1). Because we would naively expect a positive line tension, we cannot explain this.

As area conservation is assumed, it is not necessary to include a stretching term in our model. However, to verify how reliable this assumption is, we relaxed the constant area constraint (see Supporting Information for details). This necessitates the introduction of a stretching term in the free energy expression in order to account for the energy cost associated with any changes in the surface area. For simplicity, uniform stretching was assumed for the entire shell and the stretching modulus, $\kappa_s$, was related to bending modulus as we have seen above (eq.~\ref{eq:ks}). Fitting of the height histogram with the relaxed area constraint leads to a total surface area decrease for both HOPG and mica substrates. Still, the bending modulus showed negligible change. These results suggest that the contribution from stretching is minimal and that the assumption of constant surface area is valid.

Together, our findings on the orientation bias and the estimates for the contact area suggest a possible mechanism for adsorption. We have seen that the most probable orientation on HOPG is with a 3-fold axis normal to the substrate. Considering the magnitude of the radius for the contact area, pentamers should be located on its circumference, i.e. touching the substrate. 

Keeping in mind that previous indentation experiments suggest compression to occur more readily along a 3-fold than along a 5-fold axis\cite{roos2010b} and that hexameric interfaces are thought to fail more readily than pentameric ones\cite{Zandi2005, Krishnamani2016}, we propose that the main displacement upon adsorption occurs along the three-fold axis, with the hexamer at the center radially shifting its position from the surface towards the particle center, and with the stiffer pentamers acting as a stabilizing tripod. As the interfacial area grows, a point is reached where the cost of continuing the flattening of the shell is greater than the energy drop due to adhesion, at which point the virus shell is stabilized. It is worth noting that, normal mode analysis of the mechanical properties of icosahedral virus capsids\cite{Tama2005} predicts pentamers to have greater propensity to move freely. However, continuum approaches based on elastic theory predict in certain cases the opposite, i.e., pentamers being stiffer than hexamers\cite{Buenemann2008}. The latter is valid  for large ratios between elastic and bending energy contributions, for large viruses and when spontaneous curvature effects can be neglected\cite{Lidmar:2003gs}. It would be interesting to see how inclusion of substrate effects might affect these analyses. In any case, our experiments seem to support a scenario with stiffer pentamers, for a small virus.   

An issue of practical importance from a measurement perspective, is wether interactions at the substrate-virus interface affect readings of the virus stiffness in AFM indentation experiments. We have performed AFM indentation on BMV adsorbed on HOPG in SAMA buffer and plotted the elastic constants as a function of particle height. Within the framework of the proposed model, the smaller the height, the larger the virus-substrate interaction. Do particle height and elastic constants correlate? As expected, the joint histogram presented in Fig. 4 suggests that, within the experimental uncertainty, this is not the case. The particle height varied independently of the measured elastic constant $K_v$, which remained constant at $0.20\pm 0.06$ N/m. Note that, for the simple thin shell model, the elastic constant $K_v$ is proportional to the Young's modulus, which in turn is directly proportional to the bending modulus $\kappa$. To avoid inhomogeneous broadening of $K_v$ in this experiment, and keep experimental uncertainty low, we produced a nearly homogeneous BMV virion population containing mainly a subset of the viral genome ($\sim$90\% of RNA3/4) via an engineered \emph{Agrobacterium} expression system\cite{Ni2014}. Moreover, natural variation in the average radius of the virus particle (from cryo-electron microscopy measurements) is $\sim$1 nm, much smaller than the deviations measured here.  

In conclusion, we have utilized AFM imaging in liquid on flat, chemically homogeneous substrates to show that orientation and height of viruses adsorbed on a substrate depend on the virus-substrate interaction. BMV was found to adsorb preferentially with a three-fold axis parallel to the surface normal. Local deformation, measurable as a change in virus height ensues as elastic and adhesive forces equilibrate. A simple model fitting experimental data suggests that interfacial energies of tens of $k_BT$ accompany the encounter of BMV with both charged and nonpolar model substrates. As we used the simplest possible free energy  to obtain  insights into the contribution of different elastic energies,  our model is highly approximate, and it cannot reproduce the long tail in the distribution. Further investigations are required. However, local deformation at the contact area does not change the apparent elastic constant as measured by AFM indentation, which suggests that curvature elastic stress does not change upon adsorption. Since it appears that virus orientation and deformation at the surface stabilize interfacial interactions, an interesting question that might be raised is that of anisotropic deformability as yet another biologically beneficial facet of icosahedral symmetry in viruses.

\begin{acknowledgments}
This work has been supported by the U.S. Department of Energy, Office of Science, Basic Energy Sciences, under award DE-SC0010507 (to B.D., for work on atomic force microscopy) and by the Human Frontier Science Program, under award RGP0017/2012, and the National Science Foundation (DMR-1310687 to R.Z.), for modeling. The authors thank Dr. Irina Tsvetkova for her critical reading of the manuscript.
\end{acknowledgments}

\section{Figure Captions}

Figure 1. a) AFM images of BMV particles adsorbed on HOPG and mica. Several capsomers per capsid are observable at sufficient resolution to see broad morphological features of a single capsomer $\sim\ $6 nm in size. Scale bars: 10 nm. b) Orientations of the model icosahedron that correspond to virus orientations in a), respectively. c) Corresponding orientations of the BMV crystallographic model. d) and e) Distribution of surface normal directions, represented as colored lines from single measurements, with respect to an icosahedron attached to each particle and to the molecular model, on HOPG (N = 17 particles, green) and mica (N = 21 particles, blue).   

\medskip
 
Figure 2. Particle height distributions for adsorbed BMV  at equilibrium in SAMA buffer, at pH 4.5 (a), and cartoon representation of the mechanism by which particle heights may acquire a distribution when adsorbed on a surface (b).

\medskip 

Figure 3. Model fit of the particle height distribution on HOPG (a) and mica (b). At close to nominal heights ($H=1$) contact area is minimal and likelihood of desorption increased, thus populations are low. At smaller heights, adhesion comes at the cost of structural perturbation, modeled here as elastic.

\medskip 

Figure 4. Joint probability density histogram of heights and elastic constants for a homogeneous sample of BMV (N = 138) containing a genomic subset. The vertical streak aspect suggests negligible correlation between height variations due to adsorption and the magnitude of the elastic constant.

\bibliography{IndentationlessAFM}

\newpage
\begin{figure}[ht]
    \centering
    \includegraphics [width = \textwidth] {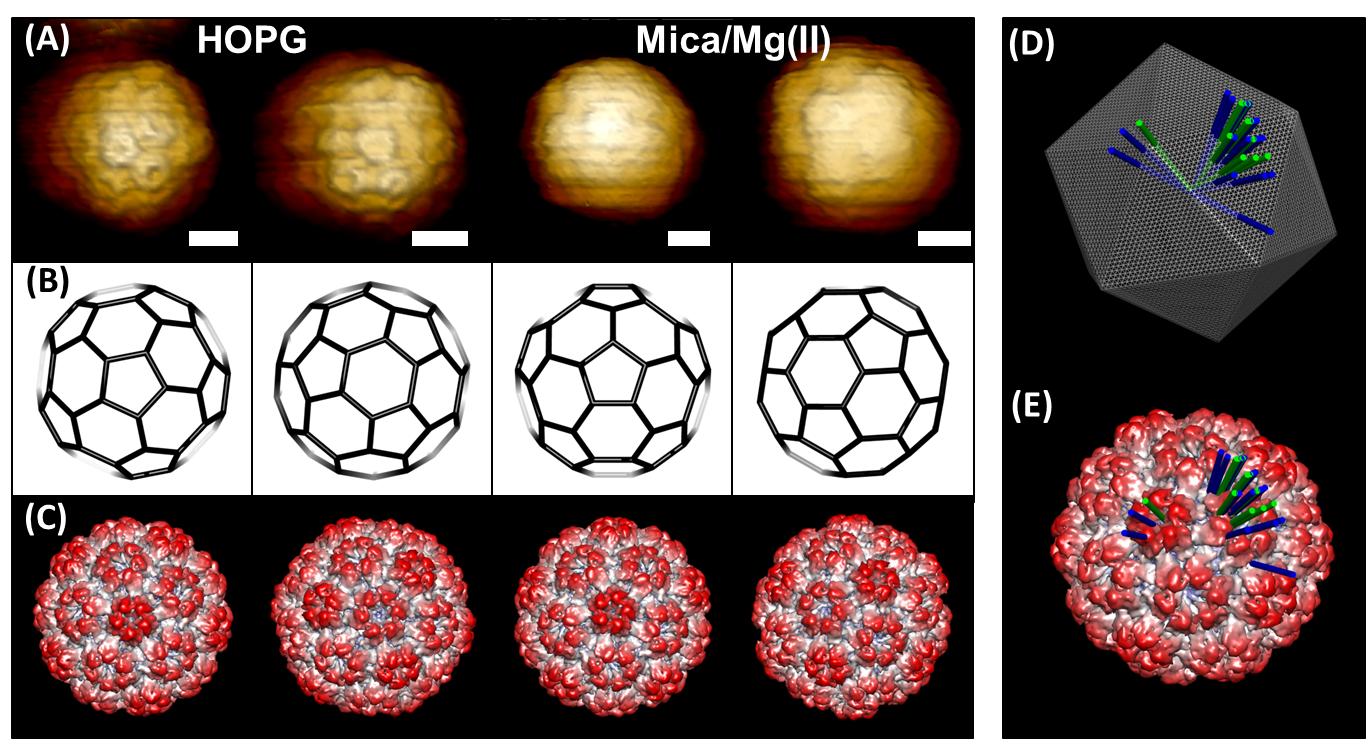}
    \caption{Figure 1}
\end{figure}

\newpage
\begin{figure}[ht]
    \centering
    \includegraphics [width = \textwidth] {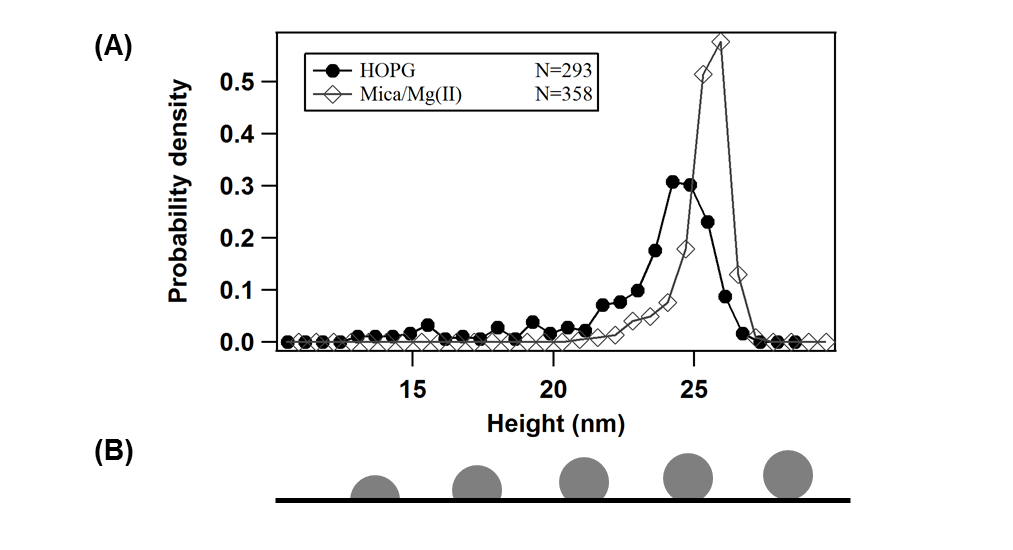}
    \caption{Figure 2}
\end{figure}

\newpage
\begin{figure}[ht]
    \centering
    \includegraphics [width = \textwidth] {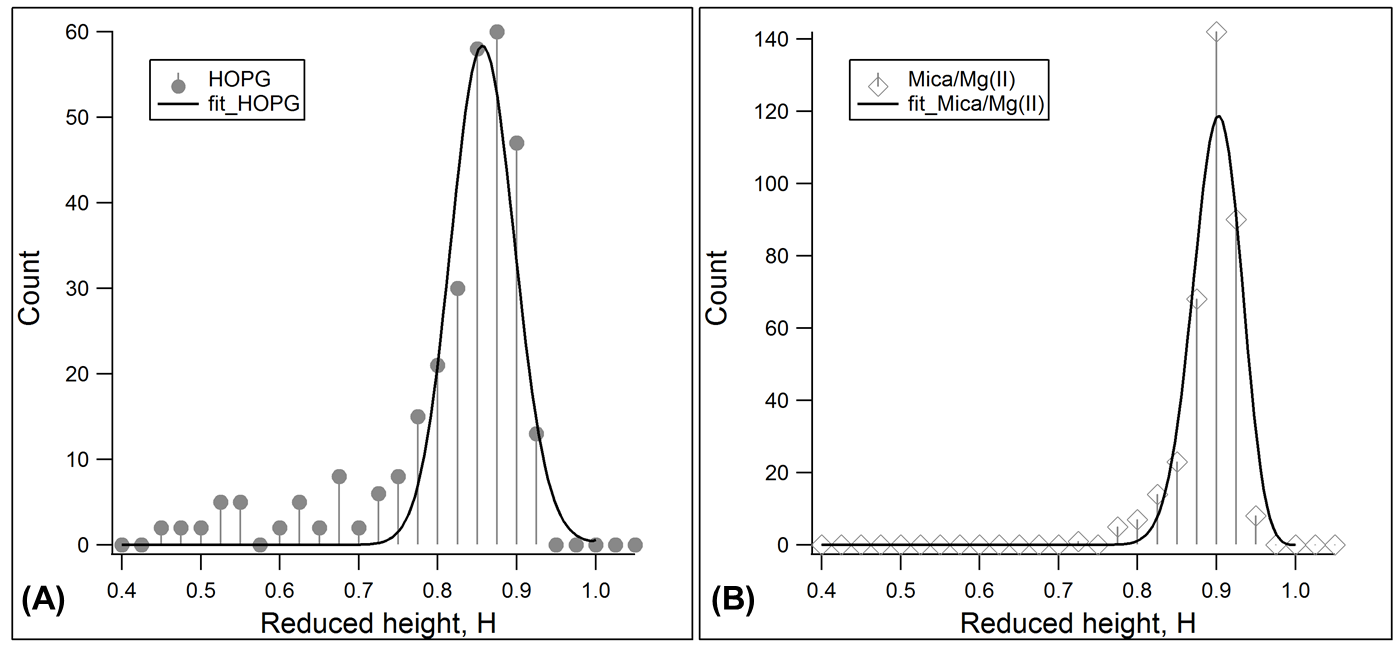}
    \caption{Figure 3}
\end{figure}

\newpage
\begin{figure}[ht]
    \centering
    \includegraphics [width = \textwidth] {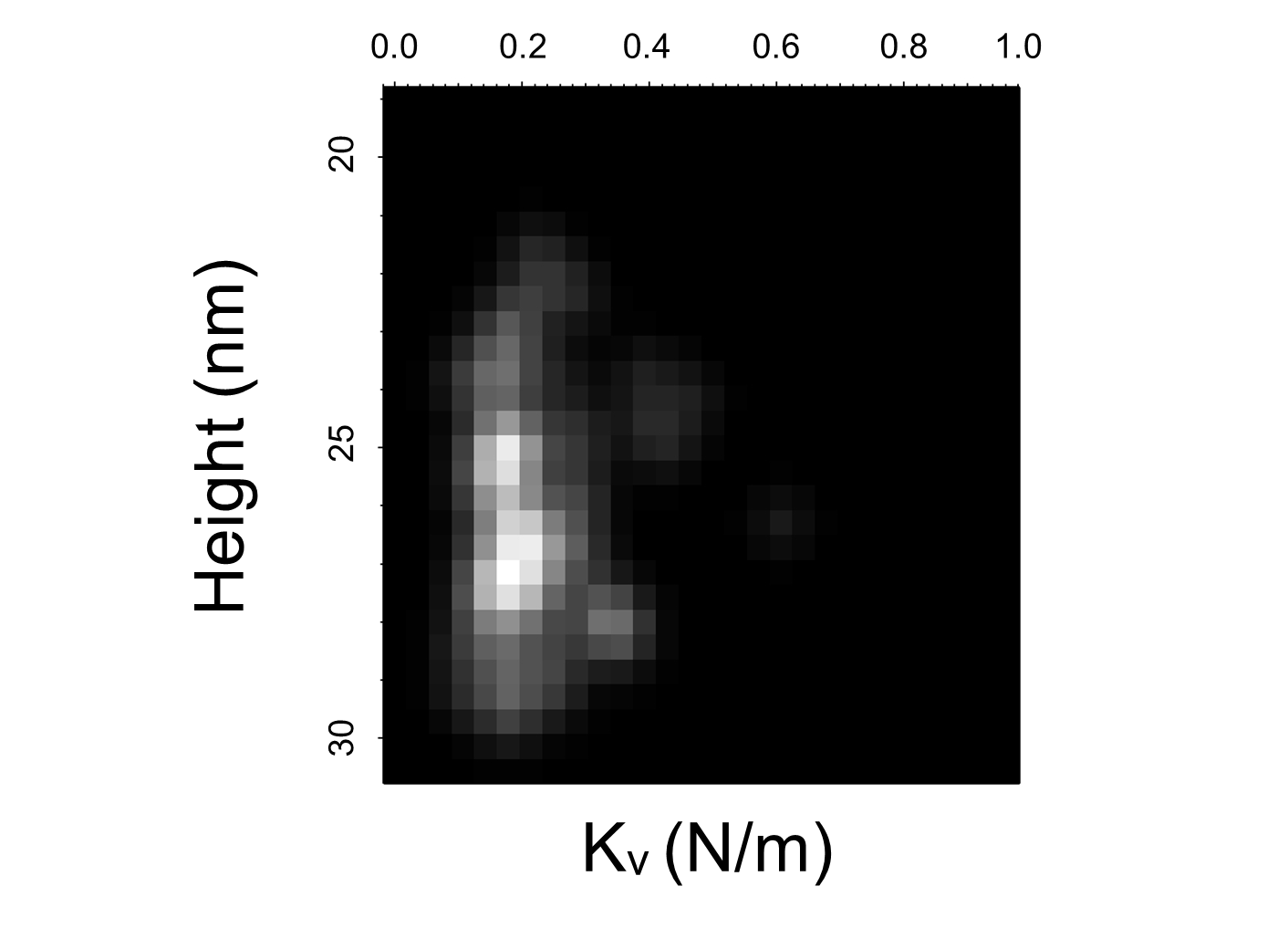}
    \caption{Figure 4}
\end{figure}

\end{document}



\title{Contact Mechanics of a Small Icosahedral Virus\\Supporting Information} 


\author{Cheng Zeng}
\affiliation{Department of Chemistry, Indiana University, Bloomington, IN 47405, U.S.A.}

\author{Mercedes Hernando-P\'{e}rez}
\affiliation{Department of Chemistry, Indiana University, Bloomington, IN 47405, U.S.A.}

\author{Xiang Ma}
\affiliation{Department of Chemistry, Idaho State University, Pocatello, ID 83209, U.S.A.}

\author{Paul van der Schoot}
\altaffiliation{Institute for Theoretical Physics, Utrecht University, Leuvenlaan 4, 3584 CE Utrecht, The Netherlands}
\affiliation{Department of Applied Physics, Eindhoven University of Technology, P.O. Box 513, 5600 MB Eindhoven, The Netherlands.}

\author{Roya Zandi}
\affiliation{Department of Physics and Astronomy, University of California at Riverside, 900 University Ave.
Riverside, CA 92521, U.S.A.}

\author{Bogdan Dragnea}
\email[]{dragnea@indiana.edu}
\affiliation{Department of Chemistry, Indiana University, Bloomington, IN 47405, U.S.A.}


\date{\today}



\pacs{}

\maketitle 

\section{Experimental Method}

\textbf{Virus production and purification.} Purification of BMV was done as reported in previous work.\cite{Peng2012} Briefly, BMV was expressed in \textit{Nicotiana benthamiana} via \textit{Agrobacterium}-mediated gene delivery. The leaves were collected seven days post infection and stored at -80 $^{\circ}$C until use. The leaves were first homogenized in virus buffer [250 mM NaOAc, 10 mM MgCl$_2$, pH 4.5] and then centrifuged at 5,000 rpm for 25 min at 5 $^{\circ}$C on a Beckman TA-10.250 rotor to remove undissolved materials. The supernatant was then layered on a 10 \% sucrose cushion (w/v) in virus buffer and centrifuged at 26,000 rpm for 3 h on a Beckman SW 32 rotor. The pellets were resuspended in 38.5 \% CsCl (w/v) in virus buffer and centrifuged at 45,000 rpm for 24 h on a Beckman Ti-70.1 rotor. The virus band was then collected and dialyzed, with three changes in 24 h, against SAMA buffer [50 mM NaOAc, 8 mM Mg(OAc)$_2$, pH 4.5]. Final purity of wild-type virus was achieved by running the virus on a Superos-6 column by FPLC. The purified virus was stored under -80 $^{\circ}$C until use.

\textbf{AFM of viral particles.} All AFM experiments were conducted with a Cypher AFM (Asylum Research, Santa Barbara, U.S.A.) in liquid. A stock solution of purified BMV was diluted in SAMA buffer (pH 4.5) into a final concentration of around 0.1 mg/mL prior each AFM experiment. A single droplet (50 $\mu$L) of diluted BMV was deposited on a newly-cleaved HOPG, mica or pretreated mica surface. Pretreated mica was prepared by immersing a newly-cleaved mica substrate in a 1M MgCl$_2$ solution for 5 min, washing with ddH$_2$O and subsequently immersing in 1M MgCl$_2$ again for 5 min. Excess solution was blotted away with filter paper. Gold-coated BioLever Mini cantilevers (Olympus, Tokyo, Japan) with a typical spring constant of 0.09 N/m and rectangular tips with 9 $\pm$ 2 nm radii of curvature were used. Tips were always prewetted with a drop (50 $\mu$L) of SAMA buffer. 
AFM images were acquired in AC mode at a constant temperature controlled by an air temperature controller. AFM images were processed with Igor Pro software. Particle height was measured by cross-sectioning image height data. Nano-indentation was performed in force-mapping mode with a typical trigger force of 700 pN. Spring constant was extracted from single force-displacement curve from the center of the particle as reported elsewhere.\cite{Roos2011}

\textbf{Orientation determination.} An icosahedral cage was first created by the hkcage function in UCSF Chimera software.\cite{Pettersen2004} This initial cage was used as a reference orientation centered in a set of spherical coordinates. For each high-resolution AFM image, the reference cage was rotated in Chimera to achieve the best match in icosahedral lattice. The orientation of the particle in this AFM image can then be represented by two spherical coordinates ($\theta$, $\phi$) of the new center point. The spherical coordinates were then used to generate final orientation maps. Note that there can be multiple points with the same orientation due to icosahedral symmetry. In this study, each particle was always represented with a single asymmetric point.
\newpage

\section{Manuscript Supporting Figures}

%
%

\begin{figure}[ht]
\includegraphics [width = 0.5 \textwidth] {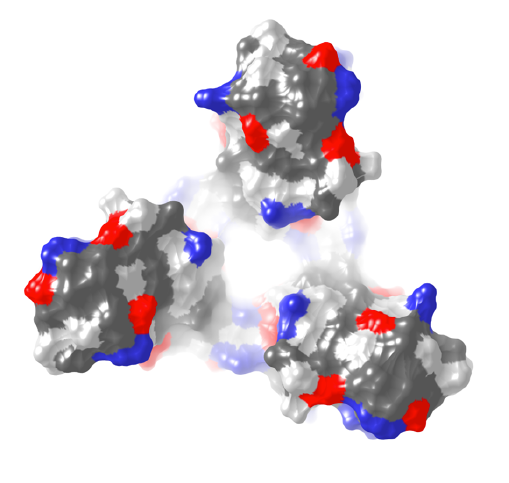}
\center
\caption{Figure SI-1. Space-filling, depth cued rendering of the asymmetric unit of the BMV coat protein oriented with the outer surface residues towards the viewer. Hydrophobic regions are dark gray, positive blue, and negative red.}
\end{figure}
\vspace{0.5 in}
\begin{figure}[ht]
\includegraphics [width = \textwidth] {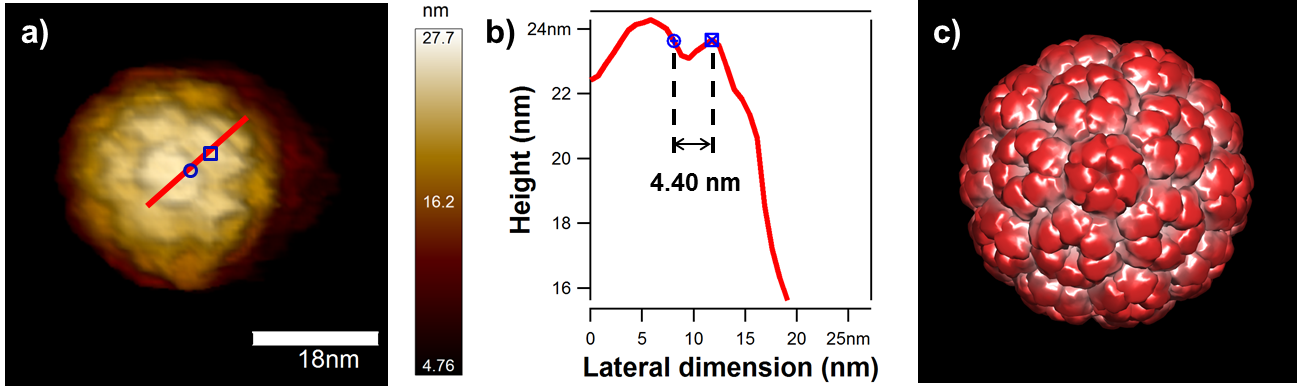}
\caption{Figure SI-2. Vertical profile along a path through an AFM height map including several capsomers; $\sim$4 nm features recognizable as part of capsomer morphology can be distinguished.}
\end{figure}

\section{Analytical Model}

To calculate the free energy of a deformed shell we make the following assumptions:
\begin{enumerate}
\item An adsorbed shell is composed of three regions: a spherical cap, a flat contact area, and a rim separating the two.
\item The total number of molecules in the particle is a constant, but the numbers in each region varies as the shell deforms.
\end{enumerate}
Unlike  the usual treatment of the adhesion of membrane vesicles\cite{Lipowsky1991}, the shape of our model (Fig. SI-3) is not parametrizable by an angle that determines the orientation of the surface normal as a function of the arclength. This is because the line of contact represents a defect or discontinuity. For viruses, this is a reasonable assumption as shown by various AFM indentation experiments and simulations, which result in an apparent discontinuity at the tip/shell contact perimeter. We assume that, at the rim, hydrophobic capsomeric contacts open and the molecules there interact with the substrate and the solvent in a manner different from the rest\cite{Israelachvili:2011p}. Each of the three regions is characterized by its own free energy density, that is, free energy per unit area. 
The total free energy is the sum of the three contributions corresponding to cap, base, and rim:
\begin{equation}
F(h) = F_c \cdot A_c + F_b \cdot A_b + F_r \cdot L_r
\end{equation}
where $F_c$ and $A_c$ are the free energy density and area of the cap; $F_b$ and $A_b$ are the free energy density and area of the base; $F_r$ and $L_r$ are the free energy density and the length of the rim. 

The probability to find a virus particle of height $h$ adsorbed on the surface (see Fig. SI-3) is determined by the Boltzmann factor:
\begin{equation}
\mathcal{P}(h) \sim e^{-(F(h)-F_0)/k_B T} = e^{-\Delta F/k_B T}\mathrm{,}
\label{eq:Bo}
\end{equation}
where $F_0$ is the free energy of an intact virus particle in solution. The prefactor of Eq. (2) is fixed by the condition of normalisation of probability, so $\int_0^{2r} P(h) = 1$. With the geometric notations of Fig. SI-3, $F(h)$ is given by: 

\begin{equation}
\label{eq:F}%
F=\frac{1}{2}\kappa \left (  \frac{2}{r}-\frac{2}{r_0} \right )^2 2 \pi r h + 2 \pi \kappa_G \frac{h}{r} -\gamma  \pi a^2 + \tau  2 \pi a\mathrm{,}
\end{equation}
where $\kappa$ is bending modulus, $\kappa_G$ is Gauss modulus,  $\gamma$ is surface adhesion energy, and $\tau$ is the rim line tension.

For an intact paricle,  $h = 2r_0$, $r = r_0$, $a =0$, eq.~\ref{eq:F} leads to:
\begin{equation}
F_0 = F(2 r_0) = 4 \pi \kappa_G\mathrm{,}  
\end{equation}
and $\Delta F$ in eq.~\ref{eq:Bo} becomes:
\begin{equation}
    \Delta F = F(h)-F_0 = \frac{1}{2}\kappa \left(  \frac{2}{r} - \frac{2}{r_0}  \right)^2 2 \pi r h  + 2 \pi \kappa_G \left(\frac{h}{r}-2\right) - \gamma \pi a^2 + 2 \pi a \tau\mathrm{,}
\label{eq:dF}
\end{equation}

Conservation of total area requires:
\begin{equation}
a = \sqrt{\frac{4r_0^2-h^2}{2}}\mathrm{,}
\label{eq:geo}
\end{equation}
which reduces the number of independent variables to two: the height after adsorption, $h$, and the initial radius, $r_0$. 
\begin{figure}[ht]
\includegraphics [width = 0.8\textwidth] {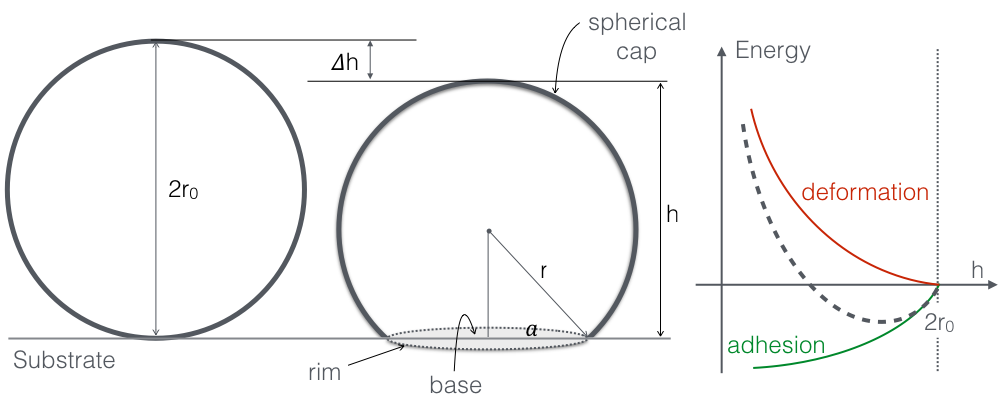}
\caption{Figure SI-3. Schematic and geometric elements notations used in the model. Upon adsorption, adhesion energy becomes more negative with a decrease in height (green) while deformation energy increases (red); total energy may develop a minimum (dotted line).}
\end{figure}

Defining the reduced height, \[ H = \frac{h}{2r_0} \] we obtain:
\begin{equation}
\begin{split}
\Delta F(H) =\ &2 \pi \kappa (1+H^2) \left[ \frac{(1-H)^2}{1+H^2} \right]^2 + 2 \pi \kappa_G \left (\frac{4H^2}{1+H^2}-2\right) -\\ &4 \pi \gamma r_0^2 \frac{1-H^2}{2} + 2 \sqrt{2} \pi \sigma d r_0 \sqrt{1-H^2}\mathrm{,} 
\end{split}
\label{eq:dFH}     
\end{equation}
where the last term corresponds to the rim and the line tension was obtained from the surface energy, $\sigma$, of exposed hydrophobic areas due to rim formation\cite{Israelachvili:2011p},  and a known shell thickness, $d\approx 4$ nm.

The interplay of the different terms in eq.~\ref{eq:dFH} is presented in Fig. SI-4.
\begin{figure}[ht]
\includegraphics [width = 0.7 \textwidth] {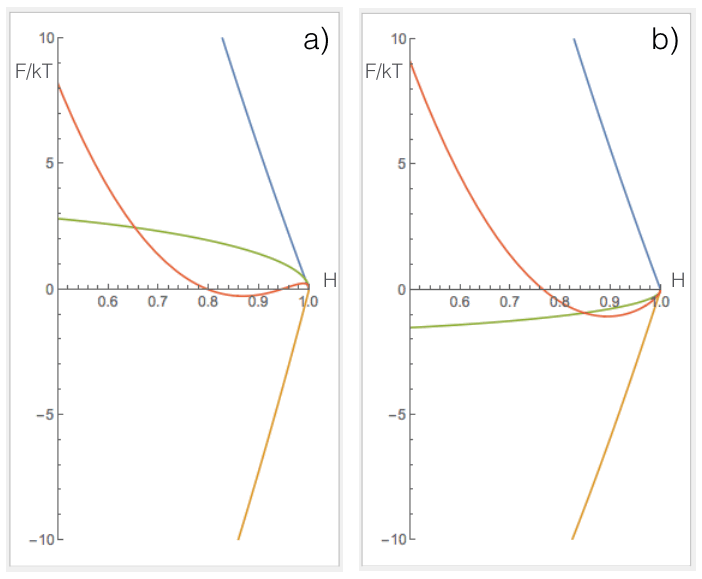}
\caption{Figure SI-4. Example of decomposition of various contributions to the total energy (red). a) A positive rim energy sets a barrier to stable adsorption. b) A negative rim energy contributes to a deeper well. Blue: Elastic energy. Orange: adhesion energy.}
\end{figure}

When the Gauss term is removed, the free energy expression becomes:
\begin{equation}
F=\frac{1}{2}\kappa \left (  \frac{2}{r}-\frac{2}{r_0} \right )^2 2 \pi r h -\gamma  \pi a^2 + \tau  2 \pi a\mathrm{,}
\end{equation}
Fitting of height histograms with this expression cannot be done unless we remove the constraint that bending modulus does not vary between substrates. Considering that the same virus and buffer conditions were used, however, the virus bending modulus should be substrate-independent. From the fittings, the bending modulus is found to be $\sim 1000$ $k_BT$ on HOPG and $\sim 4500$ $k_BT$ on mica, which are about 30 and 150 times larger than the values obtained before removal of the Gauss term. Such values are unreasonably large. Within the thin shell approximation, these values correspond to Young's moduli of 3 and 13.5 GPa, respectively, more than one order of magnitude above what is generally considered acceptable for soft RNA viruses \cite{roos2010}. Thus, it appears the Gauss term is critical in getting reasonable numbers and avoiding contradictions with established facts.

To test the assumption of surface area conservation, a stretching term was added to the free energy functional:
\begin{equation}
F=\frac{1}{2}\kappa \left (  \frac{2}{r}-\frac{2}{r_0} \right )^2 2 \pi r h + \frac{1}{2}\kappa_s \frac{(A - 4 \pi r_0^2)^2}{4 \pi r_0^2} + 2 \pi \kappa_G \frac{h}{r} -\gamma  \pi a^2 + \tau  2 \pi a
\end{equation}
where $\kappa_s$ is stretching modulus (considered the same for cap and base), and A is the total surface area. As we now allow A to vary, Eq. \ref{eq:geo} does not hold any more. It is thus necessary to express cap radius (r) and base radius (a) as a function of particle height (h) and total surface area (A).

According to geometric constraints (Fig. SI-3), there is a relationship between r, a, and h which is valid at all height values:
\begin{equation}
    r^2 = (h-r)^2 + a^2
\end{equation}
Also, total surface area (A) can be written as a summation of cap and base area:
\begin{equation}
    A = 2 \pi rh + \pi a^2
\end{equation}
Both the cap radius (r) and base radius (a) can now be written as a function of A and h:
\begin{equation}
    a = \sqrt{\frac{A - \pi h^2}{2\pi}}
\end{equation}
\begin{equation}
    r = \frac{A + \pi h^2}{4 \pi h}
\end{equation}
When there is no strain, \textit{i.e.} $h = 2r_0$, we get the initial surface area $A_0 = 4 \pi r_0^2$.
With these two relationships, the free energy can now be written as a function of reduced particle height (H) and total surface area (A):
\begin{equation}
\begin{split}
    F =\ &\kappa [A + \pi (2 r_0 H)^2][\frac{4\pi(2 r_0 H)}{A + \pi(2 r_0 H)^2} - \frac{1}{r}]^2 + \frac{1}{2}\kappa_s \frac{(A- 4\pi r_0^2)^2}{4 \pi r_0^2} +\\ &8\pi^2 (2 r_0 H)^2 \kappa_G \frac{1}{A + \pi (2 r_0 H)^2} - \frac{\gamma}{2} [ A - \pi \cdot (2 r_0 H)^2] + \tau \cdot \sqrt{2\pi [A - \pi \cdot (2 r_0 H)^2}
\end{split}
\end{equation}
$\kappa_s$ is related to the bending modulus $\kappa$ through:
\begin{equation}
    \kappa = \frac{1}{12} \kappa_s \cdot \omega^2
\end{equation}
where $\omega$ is the wall thickness.
Values of surface tension ($\gamma$) and line tension ($\tau$) obtained from the initial fitting procedure we used for the surface area and bending modulus.

For both HOPG and mica substrates, we observed $\approx 1$\% decrease in total surface area, while the bending modulus stayed virtually unchanged from the value found in the initial fitting.

\bibliography{Methods}
\bibliographystyle{apsrev4-1}

Eqs. \ref{eq:dF} and \ref{eq:Bo} were used for fiting the experimental height histograms.